\documentclass[journal,10pt,draftclsnofoot,onecolumn]{IEEEtran}
\makeatletter
\let\NAT@parse\undefined
\makeatother

\usepackage[numbers,sort&compress]{natbib}
\usepackage{times, url, graphicx, epsfig, subfigure, amsmath, amssymb}
\usepackage[ruled,linesnumbered,vlined]{algorithm2e}

\begin{document}
\title{RF-Based Charger Placement for Duty Cycle Guarantee in Battery-Free Sensor Networks (Technical Report)}
\author{Yanjun Li
        Lingkun Fu,
        Min Chen,
        Kaikai Chi 
        and Yi-hua Zhu 
\thanks{Y. Li, K. Chi and Y. Zhu are with the School of Computer Science and Technology, Zhejiang University of Technology, Hangzhou 310023, China. E-mail:
\{yjli,kkchi,yhzhu\}@zjut.edu.cn.}
\thanks{L. Fu and M. Chen are with the State Key Laboratory of Industrial Control Technology, Zhejiang University, Hangzhou 310027, China. E-mail:
lkfu@iipc.zju.edu.cn, chen\_min@zju.edu.cn.}

}
\maketitle
\thispagestyle{empty}

\begin{abstract}
Battery-free sensor networks have emerged as a promising solution to conquer the lifetime limitation of battery-powered systems. In this paper, we study a sensor network built from battery-free sensor nodes which harvest energy from radio frequency (RF) signals transmitted by RF-based chargers, e.g., radio frequency identification (RFID) readers. Due to the insufficiency of harvested energy, the sensor nodes have to work in duty cycles to harvest enough energy before turning active and performing tasks. One fundamental issue in this kind of network design is how to deploy the chargers to ensure that the battery-free nodes can maintain a designated duty cycle for continuous operation. Based on a new wireless recharge model, we formulate the charger placement problem for node's duty cycle guarantee as a constrained optimization problem. We develop both greedy and efficient heuristics for solving the problem and validate our solutions through extensive simulations. The simulation results show that the proposed particle swarm optimization (PSO)-based divide-and-conquer approach can effectively reduce the number of chargers compared with the greedy approach.
\end{abstract}


%
\IEEEpeerreviewmaketitle

\section{Introduction}
Limited lifetime has always been a major stumbling block to the applications of battery-powered sensor networks, especially to embedded sensing applications, where replacing battery is either impractical or inconvenient. To enable sustainable operation, it is desirable that sensor nodes have the capability of energy harvesting. Recently, the radio frequency (RF) energy harvesting technology has attracted significant attention due to prevalence of RF signals. As is well known, a typical application of RF energy harvesting technology is the radio frequency identification (RFID) system, where passive RFID tags receive all of their operating energy from an RFID reader and respond to the reader by reflecting the energy back. Conventional RFID tags have no capabilities of sensing and computation. However, recently University of Washington and Intel have co-designed a Wireless Identification and Sensing Platform (WISP) \cite{sample2008}, which is a sensing and computing device powered and read by RFID readers. WISPs upload sensory data to querying readers via backscatter modulation, and meanwhile they harvest energy from the RFID reader and store it in a capacitor, which powers the operation of microcontroller, data sensing, logging, and computing. More recently, tag-to-tag communication has also been realized \cite{parks2014}, which enables the formation of a battery-free sensor network.

A limitation of RF energy harvesting is the insufficiency of harvested energy compared with the energy demand. Typically, the rectified power of a WISP tag is found to be the order of $\mu$W while the power consumption of the tag in the active state is found to be the order of mW \cite{sample2008}, not to mention the power demand for performing sensing and computation tasks. Therefore, a battery-free sensor node, e.g., WISP, has to sleep and recharge for a period of time until there is enough power for it to turn active and perform tasks. Suppose the sensor nodes are duty cycled periodically to perform sensing tasks and the duty cycle is designated by the users. A fundamental problem is how to deploy minimal number of RF-based chargers, e.g., the RFID readers, so that the duty cycle of the nodes can be guaranteed.

Similar studies on RF-based charger placement have been conducted in several previous works \cite{he2013,fu2013,erol2012}. In \cite{he2013}, He \emph{et al.} consider using least number of readers to ensure that a static node placed in any position of the network receives a sufficient recharge rate for sustained operation. Their solution is inspired by the classical area coverage problem and an equilateral triangle placement pattern is proved to be optimal. In \cite{fu2013}, Fu \emph{et al.} consider another scenario where the reader is mobile and they study the optimal stop locations and the corresponding stop durations of the reader such that the total delay to charge all nodes in the network is minimized. In \cite{erol2012}, Erol-Kantarci \emph{et al.} propose to optimize the placement of RF-based chargers with the objective of maximizing the profit of user-defined missions. Generally, in the charger placement problem, the recharge model plays a key role and may greatly affect the solution. In these previous work, multi-charger recharge power is assumed to be simple summation of individual recharge power of each surrounding charger, which greatly facilitates geometric proofs of the charger placement pattern. However, few work has taken note of the fact that when the node is surrounded by multiple chargers, the signal it receives is the superposition of multiple differently delayed, attenuated, and phase-shifted signals, analog to the multipath signals \cite{patwari2011,zhang2012,yang2013}. The received power is thus not equal to summation of the received power of each individual charger.

The contributions of this paper are as follows. First, we present a new multi-charger recharge model. Second, based on the recharge model, we formulate the charger placement problem for node's duty cycle guarantee as a constrained optimization problem. Third, we develop both greedy and efficient heuristics for solving the problem and finally validate our solutions through extensive simulations.

The remainder of this paper is organized as follows. Section 2 reviews related work. Section 3 presents the charging model. In Section 4, we formulate our charger placement problem for duty cycle guarantee. In Section 5, we present our placement algorithms. We evaluate our algorithms via simulations in Section 6. Section 7 concludes this paper.

\section{Preliminaries and Problem Formulation}

\subsection{Recharge Model}

Before formulating our problem, we first present the recharge model. The recharge power of a node, denoted by $P_h$, is dependent on its received signal power, denoted by $P_r$. With a single charger, a node's received signal power can be calculated by the Friss transmission equation. The received power is then rectified and converted to electrical energy with some power loss. Hence, an empirical recharge model with single charger is as follows \cite{sample2008,he2013,fu2013}:

\begin{equation}\label{eqn:smodel}
P_h = \eta P_r=\eta \frac{{G_s}{G_r}}{{L_p}}{\left( {\frac{\lambda }{{4\pi (d + \varepsilon )}}} \right)^2}{P_s}
\end{equation}
where $\eta$ is the rectification efficiency, $G_s$ and $G_r$ are the source and receiver antenna gains respectively, $\lambda$ is the wavelength, $P_s$ is the transmit power of the charger, and $L_p$ is the polarization loss. Friis equation is useful for long distance transmission such as satellite communication, while for short distance transmission the distance $d$ should be adjusted to $d+\varepsilon$, where $\varepsilon$ is a fixed small parameter which ensures that the associated recharge power is finite in expectation. To simplify the recharge model, we leave multipath effect and antenna orientation effect out of account. In fact, multipath effect can be reduced if there is no obstacle between the energy source and the sensor node, and the antenna orientation effect can be alleviated when omni-directional antennas are used.

For a multi-charger scenario, suppose there are $K$ chargers with the same transmit power and frequency, each contributing to a differently attenuated and phase-shifted signal at node $i$. The received power from charger $k$ at node $i$ is then represented as a complex value $Z_{i,k}$, which is calculated by (\ref{eqn:mmodel1}), where $P_{_r}^{(i,k)}$ is its amplitude equal to the received power from the $k$th charger, which can be obtained via (\ref{eqn:smodel}), and $\theta_{i,k}$ is its phase denoting the corresponding time delay, measured at wavelength $\lambda$.

\begin{equation}\label{eqn:mmodel1}
Z_{i,k} = P_{_r}^{(i,k)}{e^{ - j{\theta _{i,k}}}}= \frac{{{G_s}{G_r}}}{{{L_p}}}{\left( {\frac{\lambda }{{4\pi ({d_{i,k}} + \varepsilon )}}} \right)^2}{P_s}{e^{ - j\frac{{2\pi {d_{i,k}}}}{\lambda }}}.
\end{equation}
Node $i$'s total received power $P_r^{(i)}$ is then the summation over all the $K$ components:
\begin{equation}\label{eqn:mmodel2}
P_r^{(i)} = \left\| {\sum\limits_{k = 1}^K {Z_{i,k}} } \right\|.
\end{equation}
Applying orthogonal decomposition to (\ref{eqn:mmodel2}), and further including rectification efficiency $\eta$, we obtain the multi-charger recharge model:
\begin{equation}\label{eqn:mmodel3}
P_h^{(i)} = \rho \sqrt {{{\left( {\sum\limits_{k = 1}^K {\frac{{\cos \left( {\frac{{2\pi {d_{i,k}}}}{\lambda }} \right)}}{{{{\left( {{d_{i,k}} + \varepsilon } \right)}^2}}}} } \right)}^2} + {{\left( {\sum\limits_{k = 1}^K {\frac{{\sin \left( {\frac{{2\pi {d_{i,k}}}}{\lambda }} \right)}}{{{{\left( {{d_{i,k}} + \varepsilon } \right)}^2}}}} } \right)}^2}}
\end{equation}
where $\rho  = \frac{\eta{{G_s}{G_r}{P_s}}}{{{L_p}}}{\left( {\frac{\lambda }{{4\pi }}} \right)^2}$. It is obvious that (\ref{eqn:mmodel3}) reduces to (\ref{eqn:smodel}) when $K=1$.

Different from the summation model, our proposed model captures the physical layer power features. Considering the scenario where a node is surrounded by k RF-based chargers, there are actually k major paths from the chargers (transmitters) to the node (receiver). Each path has a different length, so a wave propagating along that path takes a different amount of time to arrive at the receiver. Each path has attenuation caused by path losses, so each wave undergoes a different attenuation and phase shift. At the receiver, $k$ copies of the transmitted signal arrive, but each copy arrives at a different time delay and with a different amplitude and phase. The sum of these time delayed, scaled, and phase shifted transmitted signals is the received signal. Such a multi-path model has been verified and used in quite a few influential literatures \cite{patwari2011,zhang2012,yang2013}, which is just the basis of our proposed recharge model.

To further validate the rationality of our model, we conduct some simulations and experiments. First,we have noticed that the authors in \cite{he2013} have provided some experimental data to support their summation model, as shown in Table \ref{tab:1}. They place two readers facing each other and put a WISP tag in the middle between them. The distance between the tag and either reader varies from 0.6 to 1.2 m in increments of 0.1 m. The second row of Table \ref{tab:1} records the recharge power from reader 1 when reader 2 is turned off. The third row gives the opposite case when reader 2 is on but reader 1 is off. The fourth row gives the sum of the values in the second and third rows. The fifth row records the measured recharge power when both the readers are on, which they refer to as the joint recharge power. The last row calculates the relative errors between the sum of the individual recharge power and the joint recharge power. Therefore, to verify the performance of our proposed model, we use the data from the first and second rows of this table and calculate the joint recharge power according to our proposed model, i.e., using (2) and (3). We find that the calculation result is exactly the same as that of the summation model, i.e., the third row of Table \ref{tab:1}. This is because the distances between the tag and the two readers are the same. In such cases, the recharge powers calculated by the summation model and our proposed model are exactly the same, which can be proved by the following equation:

\begin{equation}\label{eqn:equal}
{P_r} = \left\| {{P_r}^{(1)}{e^{ - j{\theta _1}}} + {P_r}^{(2)}{e^{ - j{\theta _2}}}} \right\| = \left\| {{P_r}^{(1)}{e^{ - j\frac{{2\pi {d_1}}}{\lambda }}} + {P_r}^{(2)}{e^{ - j\frac{{2\pi {d_2}}}{\lambda }}}} \right\| = {P_r}^{(1)} + {P_r}^{(2)},
\end{equation}
where $P_r^{(1)}$ and $P_r^{(2)}$ are the individual recharge power from reader 1 and reader 2, and $\theta_1$ and $\theta_2$ are the phases of the electromagnetic waves at the tag from reader 1 and reader 2, respectively, which are decided by $d_1$ and $d_2$, i.e., the distances between the tag and the readers. When $d_1=d_2$, the above equation holds.

\begin{table*}[htbp] 
\centering
\caption{Experimental data records cited from \cite{he2013}}
\newcommand{\m}{\hphantom{$-$}}
\newcommand{\cc}[1]{\multicolumn{1}{c}{#1}}
\renewcommand{\tabcolsep}{0.45pc} 
\renewcommand{\arraystretch}{1.2} 
\begin{tabular}{@{}l|l|l|l|l|l|l|l}
\hline
  Rectified power (W)/ Distance (m)&1.2 &1.1 & 1.0 &0.9 &0.8 &0.7 &0.6\\
\hline
  Reader 1 &2.09$\times 10^{-4}$ &1.68$\times 10^{-4}$ &2.48$\times 10^{-4}$ &3.56$\times 10^{-4}$
  &3.28$\times 10^{-4}$ &6.78$\times 10^{-4}$ &4.90$\times 10^{-4}$\\
\hline
  Reader 2 &2.43$\times 10^{-4}$ &1.15$\times 10^{-4}$ &3.21$\times 10^{-4}$ &2.47$\times 10^{-4}$
  &2.37$\times 10^{-4}$ &5.01$\times 10^{-4}$ &4.43x$\times 10^{-4}$\\
\hline
  Sum of Reader 1 and Reader 2 &4.52$\times 10^{-4}$ &2.83$\times 10^{-4}$ &5.69$\times 10^{-4}$ &6.03$\times 10^{-4}$
  &5.66$\times 10^{-4}$ &11.79$\times 10^{-4}$ &9.34x$\times 10^{-4}$\\
\hline
  Reader 1 and Reader 2 &4.64$\times 10^{-4}$ &2.58$\times 10^{-4}$ &5.93$\times 10^{-4}$ &5.91$\times 10^{-4}$
  &4.95$\times 10^{-4}$ &9.98$\times 10^{-4}$ &7.74x$\times 10^{-4}$\\
\hline
  Relative error &-0.0266  &0.0891 &-0.0426 &0.0206 &0.1251 &0.1534 &0.1711\\
\hline
\end{tabular}\label{tab:1}
\end{table*}

\begin{figure}[htpb] 
  \centering
  \includegraphics[height=4cm]{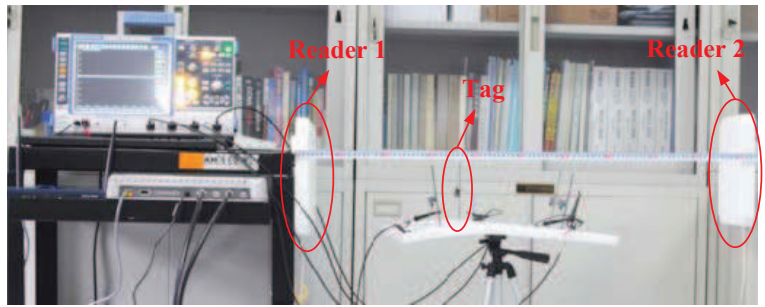}
  \caption{The experiment setup.}
  \label{fig:exp}
\end{figure}

To further verify our proposed model under other settings, we build an experiment test-bed shown in Fig. \ref{fig:exp}, which is just the same as described in \cite{he2013}, and we vary the distance settings between the tag and the readers. In the first experiment, the distance between the tag and reader 1 is fixed to 0.1m while the distance between the tag and reader 2 varies from 0.1 to 1.1 m in increments of 0.1 m. The recharge results are recorded in Table \ref{tab:2}, where the second and third rows record the recharge powers when either reader 1 or reader 2 is placed and turned on. The fourth row records the joint recharge power when both readers are on. The fifth row gives the sum of the values recorded in the second and third rows. The sixth row gives the calculated values according to our proposed model, which are obtained according to (2) and (3). Specifically,

\begin{equation}\label{eqn:equal}
{P_r} = \left\| {{P_r}^{(1)}{e^{ - j{\theta _1}}} + {P_r}^{(2)}{e^{ - j{\theta _2}}}} \right\| = \left\| {{P_r}^{(1)}{e^{ - j\frac{{2\pi {d_1}}}{\lambda }}} + {P_r}^{(2)}{e^{ - j\frac{{2\pi {d_2}}}{\lambda }}}} \right\|,
\end{equation}
where $P_r^{(1)}$ and $P_r^{(2)}$ are the individual recharge power from reader 1 and reader 2, and $\theta_1$ and $\theta_2$ are the phases of the electromagnetic waves at the tag from reader 1 and reader 2, respectively, which are decided by $d_1$ and $d_2$, i.e., the distances between the tag and the readers. In the first scenario, the distance between the tag and reader 1 ($d_1$) is fixed to 0.1m and the distance between the tag and reader 2 ($d_2$) varies from 0.1 to 1.1 m in increments of 0.1 m; the transmit frequency of readers ranges between 920-925 MHz thus the average wavelength $\lambda$ is set to 0.33m. The data in row 7 are relative errors of the summation model, which are the differences between the data in row 5 and the data in row 4. The data in row 8 are relative errors of our proposed model, which are the differences between the data in row 6 and the data in row 4.

In the second experiment, the distance between the tag and reader 1 is fixed to 0.3m and other settings are the same as the first experiment. Respective recharge results are recorded in Table \ref{tab:3}. From both Table \ref{tab:2} and Table \ref{tab:3}, we can see that, in most cases, the relative errors of our proposed model are smaller than that of the summation model. However, our proposed model still cannot perfectly match the experimental data since there are quite a few uncertain factors from the environment affecting the measurement, e.g., interferences from other equipment and small scale effects. Anyhow, we can at least show that the performance of our proposed model is not inferior to the summation model. Study on how to obtain a more accurate and practical joint recharge model remains a challenging task, yet to be further investigated.

\begin{table*}[htbp] 
\centering
\caption{Charging power of two readers, with the distance between the tag and reader 1 fixed to 0.1m}
\label{table:1}
\newcommand{\m}{\hphantom{$-$}}
\newcommand{\cc}[1]{\multicolumn{1}{c}{#1}}
\renewcommand{\tabcolsep}{0.4pc} 
\renewcommand{\arraystretch}{1.2} 
\begin{tabular}{@{}l|l|l|l|l|l|l|l|l|l|l|l}
  \hline
  Harvesting power (mW)/ Distance to reader 2 (m) &0.1 &0.2 &0.3 &0.4 &0.5 &0.6 &0.7 &0.8 &0.9 &1.0 &1.1\\
  \hline
  Reader 1 &10.13 &10.13 &10.13 &10.13 &10.13 &10.13 &10.13 &10.13 &10.13 &10.13 &10.13\\
  \hline
  Reader 2 &8.08 &4.10 &0.85 &0.16 &0.32 &0.30 &0.37 &0.37 &0.04 &0.23 &0.02\\
  \hline
  Reader 1 and reader 2&13.20 &6.20 &4.52 &2.90 &2.38 &2.52 &12.51 &1.23 &10.09 &10.27 &10.51\\
  \hline
  Sum of Reader 1 and Reader 2 &18.21 &14.23  &10.98  &10.29  &10.45  &10.43  &10.50  &10.45  &10.17  &10.36  &10.15\\
  \hline
  Result of our proposed model &18.21 &9.60 &9.47 &10.26 &10.21 &9.83 &10.29 &10.36 &10.09 &10.10 &10.15\\
  \hline
  Relative error of the summation model&5.01 &8.03 &6.46 &7.39 &8.07 &7.91 &-2.01 &9.22 &0.08 &0.09 &-0.36\\
  \hline
  Relative error of our our proposed model &5.01 &3.4 &4.95 &7.36 &7.83 &7.31 &-2.22 &9.13 &0 &-0.17 &-0.36\\
  \hline
\end{tabular}\label{tab:2}
\end{table*}

\begin{table*}[htbp] 
\centering
\caption{Charging power of two readers, with the distance between the tag and reader 1 fixed to 0.3m}
\label{table:1}
\newcommand{\m}{\hphantom{$-$}}
\newcommand{\cc}[1]{\multicolumn{1}{c}{#1}}
\renewcommand{\tabcolsep}{0.5pc} 
\renewcommand{\arraystretch}{1.2} 
\begin{tabular}{@{}l|l|l|l|l|l|l|l|l|l|l|l}
  \hline
  Harvesting power (mW)/ Distance to reader 2 (m) &0.1 &0.2 &0.3 &0.4 &0.5 &0.6 &0.7 &0.8 &0.9 &1.0 &1.1\\
  \hline
  Reader 1 &6.48 &6.48 &6.48 &6.48 &6.48 &6.48 &6.48 &6.48 &6.48 &6.48 &6.48\\
  \hline
  Reader 2 &8.35	&5.15 &4.06	&3.21 &1.92	&1.96 &0.09	&0.71 &0.33	&0.35 &0.06\\
  \hline
  Reader 1 and reader 2&9.23 &7.02	&6.77 &3.57 &2.05 &3.47 &7.78 &2.11 &1.32 &9.04 &6.40\\
  \hline
  Sum of reader 1 and reader 2&14.83 &11.63 &10.54 &9.69 &8.40 &8.44 &6.57 &7.19 &6.81 &6.83 &6.54\\
  \hline
  Result of our  proposed model &5.16 &6.83 &10.54 &6.21 &5.11 &8.19 &6.58 &5.77 &6.62 &6.74 &6.42\\
  \hline
  Relative error of the summation model&5.6 &4.61 &3.77 &6.12 &6.35 &4.97 &-1.21 &5.08 &5.49 &-2.21 &0.14\\
  \hline
  Relative error of our model&-4.07 &-0.19 &3.77 &2.64 &3.06 &4.72 &-1.2 &3.66 &5.3 &-2.3 &0.02\\
  \hline
\end{tabular}\label{tab:3}
\end{table*}


\subsection{Problem Formulation}


We define the following notations before formulating the problem.

1) Letter $A$ represent the surveillance field where total $N$ battery-free sensor nodes are located. $S=\{s_i|1\leq i\leq N\}$ represents the set of sensor nodes, where $s_i$ is the $i$th node with the coordinates of $(x_i, y_i)\in A$.

2) $C=\{c_k|1\leq k\leq K\}$ represents the charger placement, where $c_k =(x_k, y_k)\in A$ is the coordinates of the $k$th charger and $K$ is the total number of chargers. We use $|C|$ to denote the cardinality of $C$, i.e., $|C|=K$.

3) $P_a$ and $P_q$ denote the power consumptions when node is in the active and quiescent states, respectively. $\alpha$ stands for the duty cycle factor, defined as the percentage of the time during which the node is active. Hence the required recharge power for sustainable operation is $P_{req}(\alpha)=\alpha P_a+(1-\alpha)P_q$.


Our charger placement problem is formulated as follows.

\textbf{Minimal Charger Placement Problem (MCPP)}: \emph{Given a surveillance field $A$ and a set of battery-free sensor nodes $S$, find charger placement $C$ such that the number of chargers $|C|$ is minimized, subject to the energy harvesting constraint: $\forall i,P_h^{(i)} \ge P_{req}(\alpha)$.}

MCPP is a non-linear and non-convex optimization problem. We give solutions in the following section.

\section{Solutions}

In the following, we propose three approaches for solving the optimization problem.

\subsection{Greedy Approach}

The surveillance field is divided into $X\times Y$ grids. Potential coordinates of a charger is supposed to be in the center of each grid, called the grid point. The placement strategy can thus be represented by a $X\times Y$ matrix $D$, in which $D(x,y)$ denotes the number of chargers placed at grid point $(x,y)$. Once we obtain the matrix $D$, the charger placement $C$ can be easily derived. Typically, $D(x,y)$ is either 0 or 1. However, in very rare cases when the duty cycle requirement is so high that more than one charger need to be placed in a small neighborhood, it is possible that $D(x,y)$ takes an integer larger than 1. In each iteration, the charger is placed at the grid point where the number of sensor nodes satisfying the energy harvesting constraint is maximized. The details are shown in Algorithm 1. It is worth mentioning that the granularity of the grid is adjusted based on
the precision requirement and computation cost one can afford.

{\renewcommand\baselinestretch{1}\selectfont
\begin{algorithm}[htpb]
\KwIn{surveillance field $A$, the set of sensor nodes $S$}
\KwOut{$D$}
\DontPrintSemicolon \SetAlgoLined
$D=zeros(X, Y)$;$S'=\varnothing$;\\
\While {$|S'|<N$}
{find a grid point $(x,y)$ that maximizes $|S'|$ with $S' = \{ {s_i}|P_h^{(i)} \ge P_{req}(\alpha),i = 1,...,N\} $;\\
$D(x, y)=D(x, y)+1$;}
\caption{Greedy approach}
\end{algorithm}\label{algo:greedy}
\par}

\subsection{PSO-Based Approach}

As greedy algorithm is likely to get trapped in local optima, we propose Particle Swarm Optimization (PSO) based approach to solve MCPP. PSO is a population-based stochastic searching algorithm inspired by social behavior of bird flocking, animal hording, or fish schooling \cite{kennedy2010}. It works with a group of ``particles''. Each particle has a position vector and a velocity vector. The position vector simulates a candidate solution to the optimization problem, and the velocity vector denotes the position-changing tendency. For MCPP, We define the position vector of particle $i$ as the the 2-D coordinates of $k$ chargers, i.e., $\textbf{x}_i=\{x_1,y_1,x_2,y_2,...,x_k,y_k\}$, which has dimension of $2k$. To search for the optimal solution, a particle iteratively updates its velocity and current position according to (\ref{eqn:pso1}) and (\ref{eqn:pso2}).
\begin{equation}\label{eqn:pso1}
\textbf{v}_i(t+1)=w\cdot \textbf{v}_i(t)+\varphi_p\cdot \textbf{r}_p\cdot(\textbf{p}_i-\textbf{x}_i(t))+\varphi_g\cdot \textbf{r}_g\cdot(\textbf{p}_g-\textbf{x}_i(t))
\end{equation}
\begin{equation}\label{eqn:pso2}
\textbf{x}_i(t+1)=\textbf{x}_i(t)+\textbf{v}_i(t+1).
\end{equation}
Here, $\textbf{x}_i(t)$ and $\textbf{v}_i(t)$ are the current position and velocity of particle $i$, respectively; $\textbf{p}_i$ is the particle's best known position; $\textbf{p}_g$ is the swarm's best position; $\textbf{r}_p$ and $\textbf{r}_g$ are two random vectors in $U(0, 1)$; $w$, $\varphi_p$ and $\varphi_g$ are constants selected in order to control the efficacy of PSO algorithm. The update process is repeated for a fixed number of iterations.
We search for the optimal charger placement with incremental number of chargers. The details are shown in Algorithm 2. It begins with $k = 1$ and iterates for incremental $k$. In each iteration, the number of sensor nodes satisfying the constraint $P_h^{(i)} \ge {\alpha}P_a+(1-\alpha)P_q$ is maximized. Once the constraint is satisfied for all nodes, the optimal solution is found.

{\renewcommand\baselinestretch{0.9}\selectfont
\begin{algorithm}[htpb]
\KwIn{surveillance field $A$, the set of sensor nodes $S$}
\KwOut{charger placement $C$ where $|C|$ is minimized}
\DontPrintSemicolon \SetAlgoLined
$k=0$;$S'=\varnothing$;\\
\While {$|S'|<N$}
{$k=k+1$;\\
use the PSO solver to find $k$ charger locations in $A$ that maximize $|S'|$ with $S' = \{ {s_i}|P_h^{(i)} \ge P_{req}(\alpha),i = 1,...,N\} $;
}
\Return $C$
\caption{PSO-based approach}
\end{algorithm}\label{algo:pso}
\par}

\subsection{PSO-Based Divide-and-Conquer Approach}

A weakness of PSO is that in high dimensional solution space, it is hard to reach optima in each dimension, resulting in a low optimizing precision or even failure. Hence for a large-scale surveillance field or a high power demand, the PSO-based approach may be inefficient. A straightforward solution is to use a divide-and-conquer (D\&C) approach, which divides the nodes into a number of small clusters, recharges the clusters one by one using the PSO solver and then combines all local solutions into a global solution. However, a key challenge in implementing this approach is that the local problems (i.e., charger placement for individual cluster) are dependent. This is because a charger may contribute to multiple nodes in different clusters. As a result, solving the local problems separately without considering the interdependence between local solutions may lead to an inefficient global solution.

Due to the spatial decay of signal power, the chargers far away from the nodes makes little contribution to the recharge power.
Therefore, for any node to be charged, we define the \emph{contributive recharge region}(\emph{c-region}), as the disc of radius
$R$ centered at the node. The radius $R$ is referred to as \emph{contributive recharge radius} (\emph{c-radius}), which is defined as the distance $d$ in the solution to the equation $P_h=\delta P_{req}(\alpha)$, i.e., $R=\sqrt {\frac {\rho}{\delta P_{req}(\alpha)}}-\varepsilon$, where $\delta \in (0,1)$ is a tuning parameter related to node density. The optimal c-radius is therefore dependent on the recharge demand and network density and should be carefully chosen. On one hand, a conservative c-radius confines charger's recharge capability though they may contribute to the nodes outside the c-radius. On the other hand, a large c-radius may defeat the purpose of clustering and result in poor optimization performance.


We now describe the basic idea of our PSO-based divide-and-conquer (PSO-D\&C) approach. In the divide stage, the nodes to be charged are grouped into clusters according to their proximity. We adopt a greedy clustering algorithm called Quality Threshold (QT) algorithm \cite{heyer1999}, whose objective is to find minimal number of clusters that group the sensor nodes in geographical proximity. Specifically, in each iteration, a candidate cluster is created centering at each unclustered node $s$ with the nodes within the c-region of $s$ as members. The candidate with the most members is kept as a real cluster. The clustering procedure continues until there is no unclustered node left. Suppose QT yields $M$ clusters. Denote $A_m$ as the c-region of cluster $m$ with cluster head $s_{m,1}$, and $S_m$ as the set of nodes in cluster $m$, $S_m=\{s_{m,i}|i=1,2,...,|S_m|\}$. In the conquer stage, the PSO solver is executed to find an optimal local solution for each cluster. The advantage of Algorithm 3 is that the chargers placed in previous clusters can be reused by the current cluster. Hence the charger placement is jointly optimized among adjacent clusters, which effectively reduces the total number of chargers. 

{\renewcommand\baselinestretch{0.9}\selectfont
\begin{algorithm}[htpb]
\KwIn{$A$, $S$, $M$ clusters}
\KwOut{charger placement $C$ where $|C|$ is minimized}
\DontPrintSemicolon \SetAlgoLined
$C=\varnothing$;\\
\For {$m=1:M$} {
$k=0$;$S'=\varnothing$;\\
\While {$|S'|<|S_m|$}
{$k=k+1$;\\
use the PSO solver to find $k$ additional charger locations in $A_m$, denoted by $\Delta$, that maximize $|S'|$ with $S' = \{ {s_{m,i}}|P_h^{(m,i)} \ge P_{req}(\alpha),i = 1,...,|S_m|\} $;}
$C=C\cup \Delta$;\\}
\caption{PSO-D\&C approach}
\end{algorithm}\label{algo:dc_pso}
\par}

\section{Performance Evaluation}

In this section, simulations are conducted with different sensor node layouts to evaluate the proposed greedy and PSO-D\&C approaches.
Considering pure PSO-based approach only works efficiently when the network is small in scale or the recharge demand is extremely low, which can be regarded as a special case of PSO-D\&C approach, we do not evaluate its performance separately.
The parameters related to the recharge model are set according to real hardware measurements of the node, i.e., WISP4.1DL
and the RFID reader, Impinj Octane3 Speedway \cite{sample2008,he2013}. That is $\eta=0.3$, $G_s=8$ dBi, $G_r=2$ dBi, $L_p=3$ dB, $\lambda=0.33$ m, $P_s=1$ W, and $\varepsilon=0.2316$ m.
For WISP node, the average current consumptions in active and quiescent states are 600 $\mu$A and 1 $\mu$A, respectively, while the operation voltage is 1.8 V, thus the average power consumptions in active and quiescent states are $P_a = 1.08\times 10^{-3}$ W and $P_q = 1.8\times 10^{-6}$ W, respectively.
In each simulation scenario, WISP nodes are placed regularly or randomly in a $12\times 12$ m$^2$ square area. In the greedy approach, the grid size is set to 0.1 m. In the PSO-D\&C approach, the parameters required for PSO follow the routine settings given in \cite{kennedy2010}.

\begin{figure}[htpb]
 \subfigure[Greedy, $N=144$, $K=28$]{
    \label{sfig:gd_grid} 
    \begin{minipage}[b]{0.45\textwidth}
     \centering
      \includegraphics[width=2.5in]{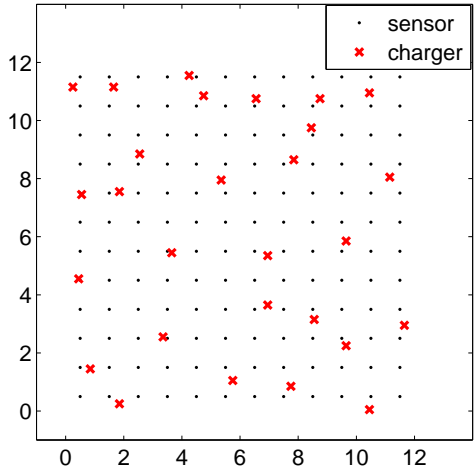}
    \end{minipage}}
  \subfigure[PSO-D\&C, $N=144$, $K=26$]{
    \label{sfig:dc_grid} 
    \begin{minipage}[b]{0.45\textwidth}
     \centering
      \includegraphics[width=2.5in]{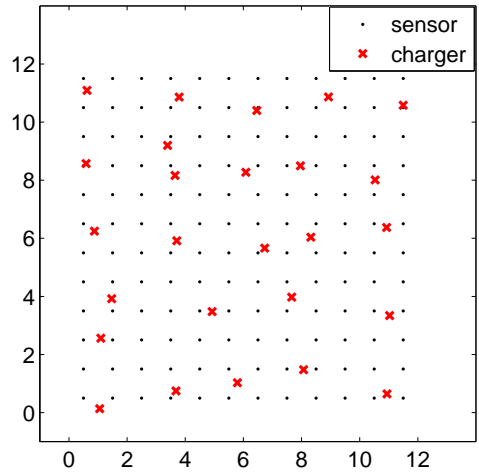}
    \end{minipage}}\\
  \subfigure[Greedy, $N=120$, $K=18$]{
    \label{sfig:gd_rand} 
    \begin{minipage}[b]{0.45\textwidth}
    \centering
      \includegraphics[width=2.5in]{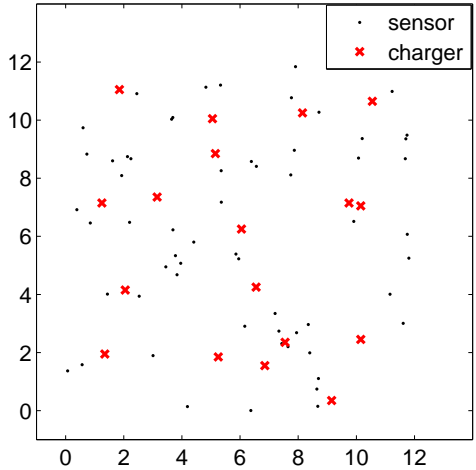}
    \end{minipage}}
  \subfigure[PSO-D\&C, $N=120$, $K=14$]{
    \label{sfig:dc_rand} 
    \begin{minipage}[b]{0.45\textwidth}
     \centering
      \includegraphics[width=2.5in]{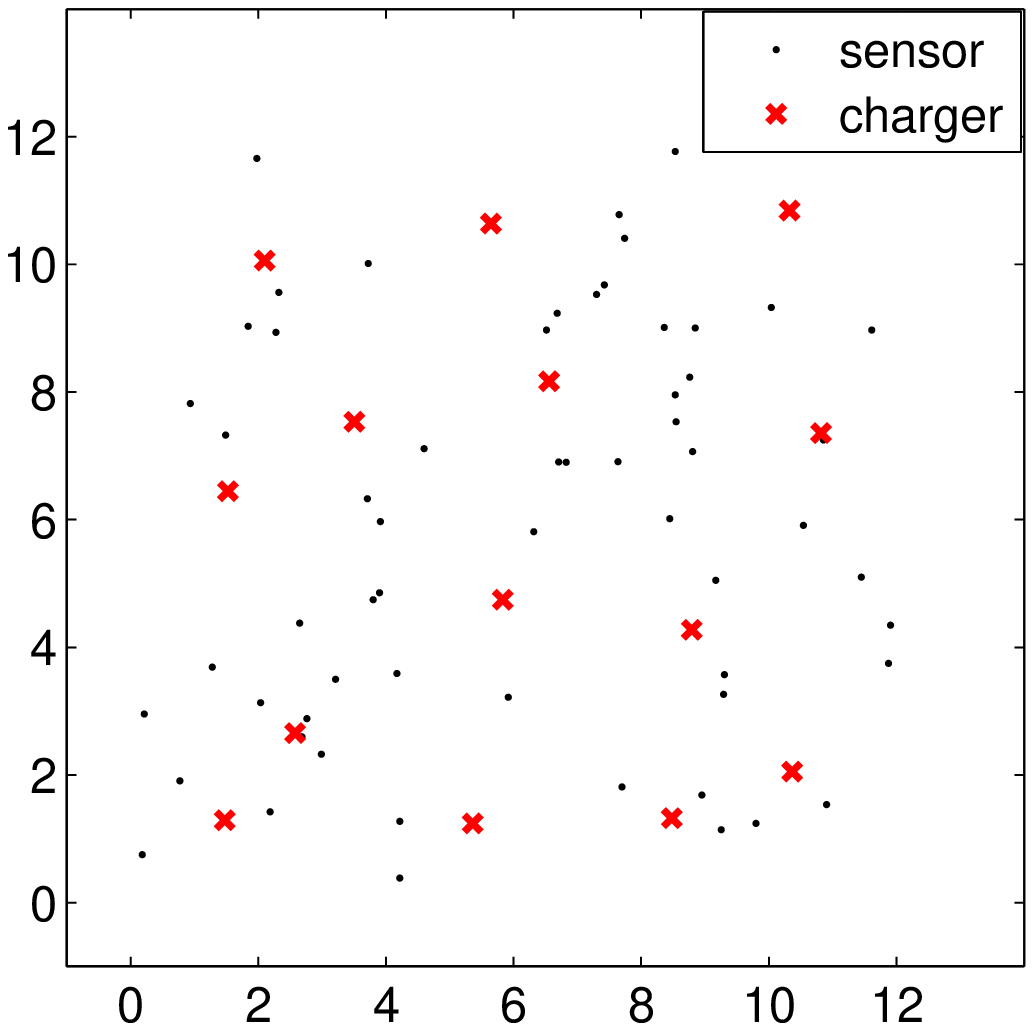}
    \end{minipage}}
      \caption{Charger placements using the greedy and PSO-D\&C approaches, for (a) and (b), $\alpha=0.5$, for (c) and (d), $\alpha=0.3$. }
  \label{fig:ts} 
\end{figure}

\begin{figure}[t]
\centering
 \subfigure[Regular, $N=144$]{
    \label{sfig:rg144} 
    \begin{minipage}[b]{0.45\textwidth}
     \centering
      \includegraphics[width=2.5in]{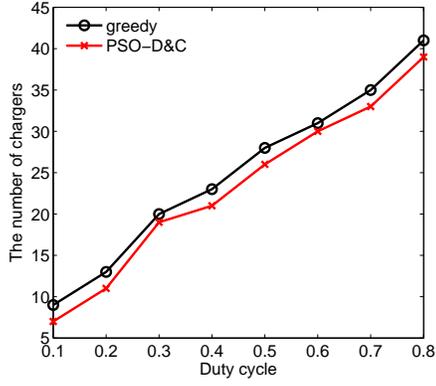}
    \end{minipage}}
  \subfigure[Random, $N=120$]{
    \label{sfig:rand120} 
    \begin{minipage}[b]{0.45\textwidth}
     \centering
      \includegraphics[width=2.5in]{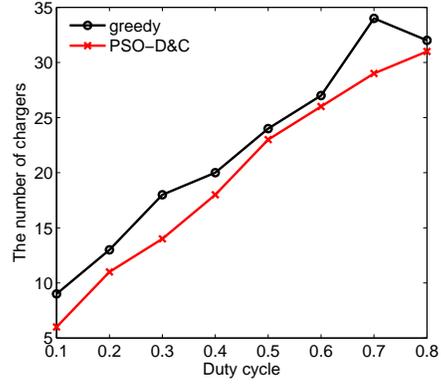}
    \end{minipage}}\\
  \subfigure[Regular, $N=64$]{
    \label{sfig:rg64} 
    \begin{minipage}[b]{0.45\textwidth}
     \centering
      \includegraphics[width=2.5in]{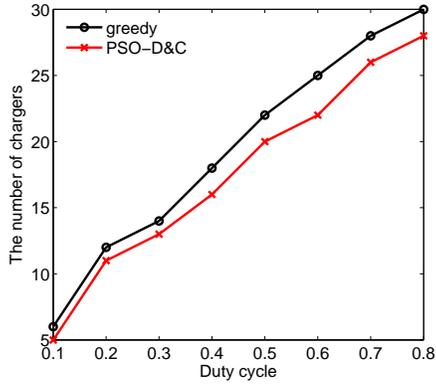}
    \end{minipage}}
  \subfigure[Random, $N=60$]{
    \label{sfig:rand60} 
    \begin{minipage}[b]{0.45\textwidth}
     \centering
      \includegraphics[width=2.5in]{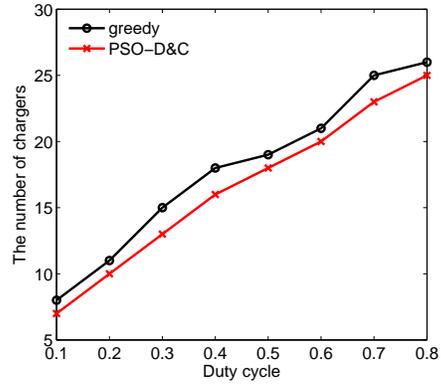}
    \end{minipage}}
      \caption{The number of required chargers with increasing duty cycle values under different network scenarios. }
  \label{fig:ts} 
\end{figure}

In the first set of simulations, total 144 WISP nodes are regularly distributed, as shown in Fig. \ref{sfig:gd_grid} and Fig. \ref{sfig:dc_grid}. Fig. \ref{sfig:rg144} shows the number of required chargers computed by the greedy and PSO-D\&C algorithms with the duty cycle requirement $\alpha$ increases from 0.1 to 0.8. The charger placements are shown in Fig. \ref{sfig:gd_grid} and Fig. \ref{sfig:dc_grid} with $\alpha$ fixed at 0.5. In the second set of simulations, total 120 sensor nodes are randomly scattered in the field, as shown in Fig. \ref{sfig:gd_rand} and Fig. \ref{sfig:dc_rand}. Fig. \ref{sfig:rand120} shows the results under different duty cycle values. The charger placements are also shown in Fig. \ref{sfig:gd_rand} and Fig. \ref{sfig:dc_rand} with $\alpha$ fixed at 0.3. We also evaluate the algorithm performances with 64 regularly distributed nodes and 60 randomly distributed nodes. The number of required chargers computed by the two algorithms are shown in Fig. \ref{sfig:rg64} and Fig. \ref{sfig:rand60}. In summary, we have the observation that, in all simulation scenarios, the PSO-D\&C algorithm consistently outperforms the greedy algorithm. The average performance gain is around 6\%.

The most related work to our work is \cite{he2013}. However, the major objectives of their work and ours are different. One of their objectives is to use least number of readers to ensure that a static tag placed in any position of the network will receive a sufficient recharge rate for sustained operation. Therefore, their problem is analog to area coverage problem. While in our work, the positions of the tags are known beforehand. Our objective is to use least number of readers to ensure that the tags at given locations can maintain a designated duty cycle for continuous operation. So our problem is more analog to point coverage problem. Generally, if using the same recharge model, the solution to their problem is obviously the solution to our problem but not the optimal one because area coverage requires more nodes than coverage of specific points if using the same coverage model. This is one of the reasons that we didn¡¯t compare the performance of our solution with that of the solution in \cite{he2013}. Another reason is that under our proposed recharge model, considering phase offset, the joint recharge powers at some location points are less than that computed by the summation model. Therefore, the reader (charger) deployment pattern proposed in \cite{he2013} cannot guarantee the required performance under our proposed model. It is thus unfair to compare the performances because their solution may require less number of readers but cannot satisfy the system requirement of node¡¯s sustainable operation.

\begin{figure}[htpb]
 \subfigure[Deployment pattern under summation model]{
    \label{sfig:r3} 
    \begin{minipage}[b]{0.45\textwidth}
     \centering
      \includegraphics[width=2in]{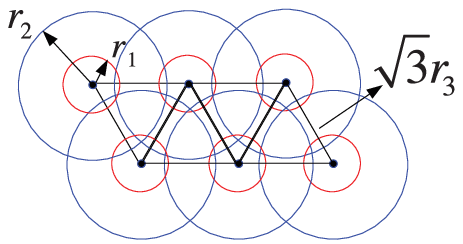}
    \end{minipage}}
  \subfigure[Deployment pattern under disk model]{
    \label{sfig:r1} 
    \begin{minipage}[b]{0.45\textwidth}
     \centering
      \includegraphics[width=2in]{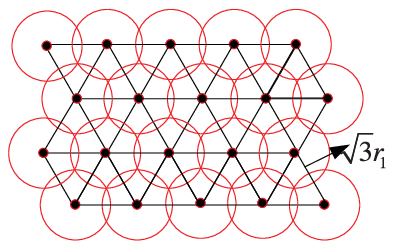}
    \end{minipage}}
      \caption{ Illustration of the deployment pattern cited from \cite{he2013}. }
  \label{fig:r} 
\end{figure}

In the following, we evaluate the performance of the deployment pattern mentioned in \cite{he2013} by our proposed new recharge model. Fig. \ref{sfig:r3} shows the deployment pattern proposed in \cite{he2013} under summation model. Fig. \ref{sfig:r1} is the deployment pattern under traditional disk model. $r_1$ and $r_3$ in Fig. \ref{fig:r} are calculated by the following equations:

\begin{equation}\label{eqn:r}
{r_1} = \sqrt {\frac{\rho }{{{P_{req}}(\alpha )}}}  - \varepsilon;~~{r_3} = \sqrt {\frac{{3\rho }}{{{P_{req}}(\alpha )}}}  - \varepsilon
\end{equation}

\begin{figure}[htpb]
 \subfigure[Required number of chargers]{
    \label{sfig:place} 
    \begin{minipage}[b]{0.45\textwidth}
     \centering
      \includegraphics[width=2.5in]{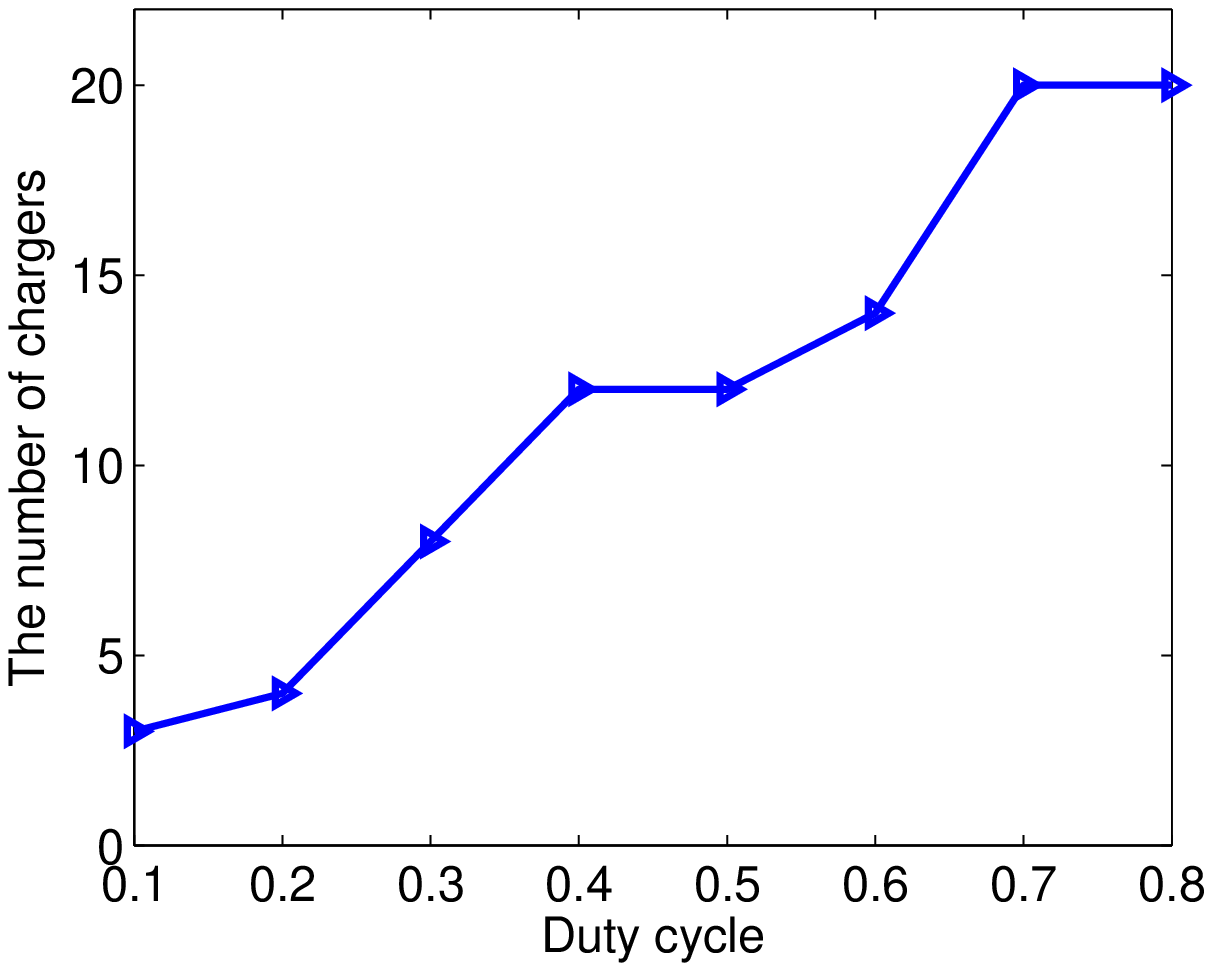}
    \end{minipage}}
  \subfigure[Sustainable node ratio]{
    \label{sfig:cov} 
    \begin{minipage}[b]{0.45\textwidth}
     \centering
      \includegraphics[width=2.5in]{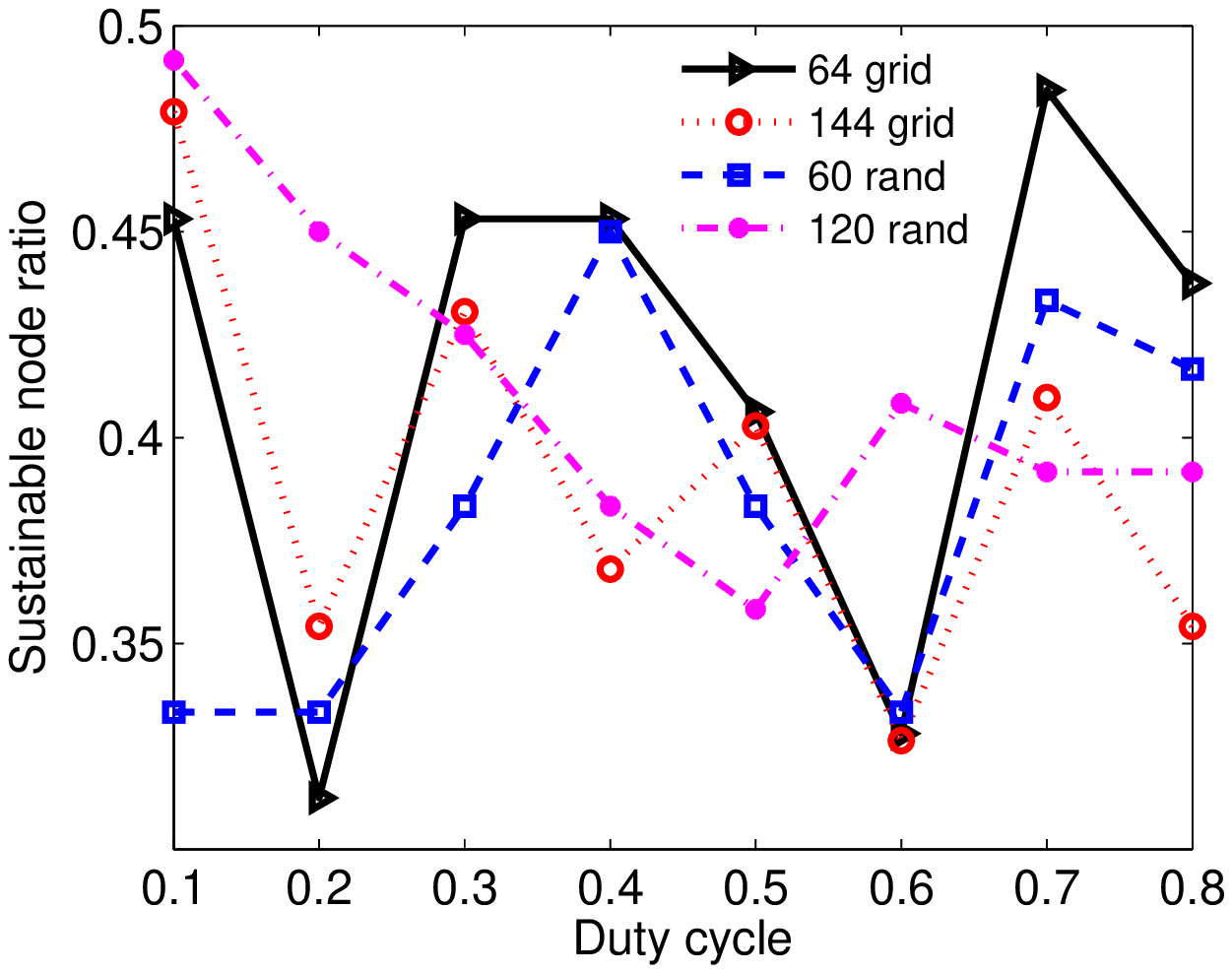}
    \end{minipage}}
    \caption{Performances under summation model with increasing duty cycle. }
      \label{fig:he} 
\end{figure}

With our simulation settings, the number of required chargers under summation model is shown in Fig. \ref{sfig:place}. Although less number of chargers is needed, the system requirement cannot be satisfied if evaluated by our proposed recharge model. We can see from Fig. \ref{sfig:cov}, the sustainable node ratio is extremely low (below 50\% in all situations).

\begin{figure}[htpb]
  \subfigure[Required number of chargers]{
    \label{sfig:dense_place} 
    \begin{minipage}[b]{0.45\textwidth}
     \centering
      \includegraphics[width=2.5in]{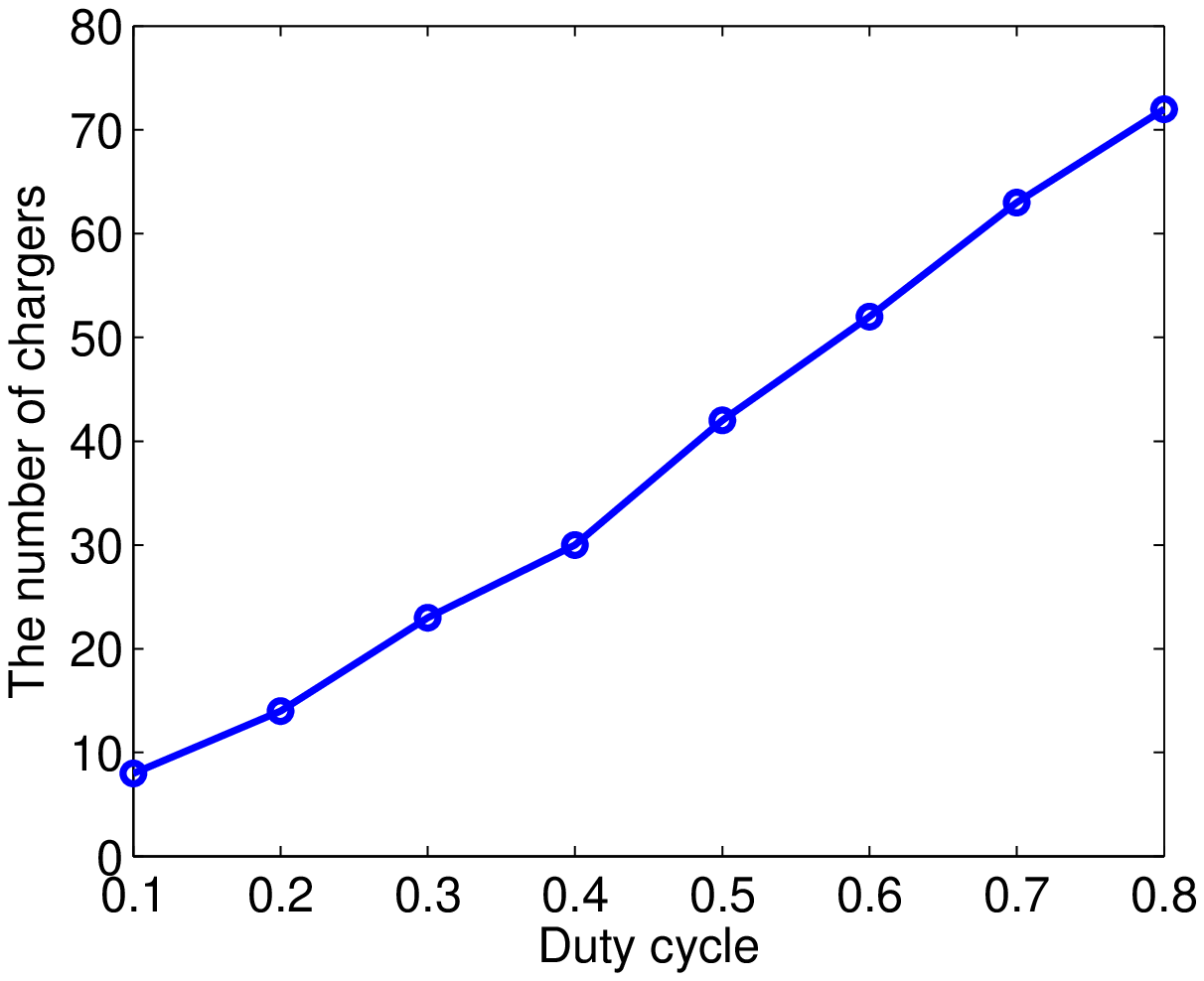}
    \end{minipage}}
  \subfigure[Sustainable node ratio]{
    \label{sfig:dense_cov} 
    \begin{minipage}[b]{0.45\textwidth}
     \centering
      \includegraphics[width=2.5in]{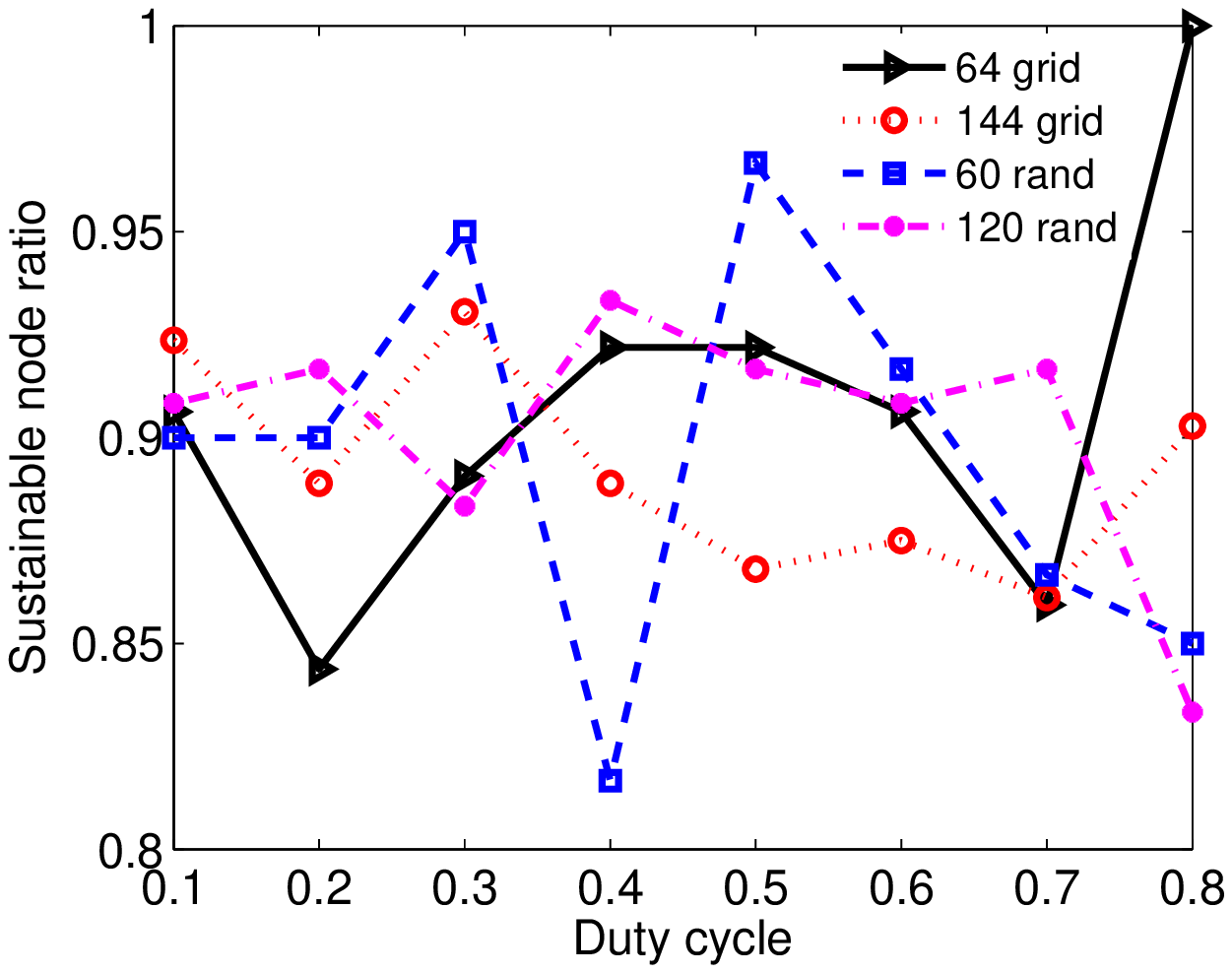}
    \end{minipage}}
      \caption{Performances under traditional disk model with increasing duty cycle. }
  \label{fig:he_dense} 
\end{figure}

Similarly, the number of required chargers under traditional disk model is shown in Fig. \ref{sfig:dense_place}. Much more number of chargers is needed. However, the system requirement still cannot be fully satisfied if evaluated by our proposed recharge model. We can see from Fig. \ref{sfig:dense_cov}, the sustainable node ratio cannot reach 100\% in most situations.

Furthermore, to date we haven¡¯t found any other paper investigating such similar RF-based charger placement issue. So we only give the simulation results of our proposed PSO-based divide and conquer approach and the greedy approach.


\section{Conclusion}
In this paper, we first present a new recharge model considering multi-signal superposition for RF-based battery-free sensor networks. Based on the recharge model, we study the problem of how to deploy minimal number of chargers to guarantee the duty cycle of battery-free sensor nodes. Both greedy and efficient PSO-based heuristics are proposed to solve the problem. The derived solutions are validated through extensive simulations that indicate the proposed PSO-D\&C approach effectively reduces the number of chargers compared with the greedy approach.

\ifCLASSOPTIONcaptionsoff
  \newpage
\fi

\bibliographystyle{IEEEtran}
\bibliography{refs}
\end{document}